\documentclass[a4paper,11pt]{article}
\usepackage{pos} 

\title{Impact of supersymmetry on the dynamical emergence of the spacetime in the type IIB matrix model with the Lorentz symmetry ``gauge fixed''\footnote{Preprint number: KEK-TH-2806}}
\ShortTitle{IIB matrix model with the Lorentz symmetry "gauge fixed"}

\author[a]{Konstantinos N. Anagnostopoulos}
\author*[b]{Takehiro Azuma}
\author[c,d]{Mitsuaki Hirasawa}
\author[e,f]{Jun Nishimura}
\author[g]{Asato Tsuchiya}
\author[e]{Naoyuki Yamamori}

\affiliation[a]{Physics Department, School of Applied Mathematical and Physical Sciences, National Technical University, Zografou Campus, GR-15780 Athens, Greece}
\affiliation[b]{Institute for Fundamental Sciences, Setsunan University,\\ 17-8 Ikeda Nakamachi, Neyagawa, Osaka, 572-8508, Japan}
\affiliation[c]{Department of Physics, University of Milano-Bicocca, Piazza della Scienza 3, I-20126 Milano, Italy}
\affiliation[d]{Sezione di Milano Bicocca, Istituto Nazionale di Fisica Nucleare,\\ Piazza della Scienza, 3, I-20126 Milano, Italy}
\affiliation[e]{KEK Theory Center, High Energy Accelerator Research Organization,\\ 1-1 Oho, Tsukuba, Ibaraki 305-0801, Japan}
\affiliation[f]{Graduate University for Advanced Studies (SOKENDAI), 1-1 Oho, Tsukuba, Ibaraki 305-0801, Japan}
\affiliation[g]{Department of Physics, Shizuoka University, 836 Ohya, Suruga-ku, Shizuoka 422-8529, Japan}

\emailAdd{konstant@mail.ntua.gr}
\emailAdd{azuma@mpg.setsunan.ac.jp}
\emailAdd{Mitsuaki.Hirasawa@mib.infn.it}
\emailAdd{jnishi@post.kek.jp}
\emailAdd{tsuchiya.asato@shizuoka.ac.jp}
\emailAdd{yamamori@post.kek.jp}

\abstract{The type IIB matrix model has been proposed as a nonperturbative formulation of superstring theory. While numerical simulations of this model are essential for probing nonperturbative effects, such as the emergence of time and an expanding 3--dimensional space, they are hindered by the sign problem. We address this using the Complex Langevin Method (CLM). Furthermore, to suppress spurious numerical artifacts that originate from large Lorentz boosts due to the Lorentz symmetry of the model, we nonperturbatively fix the Lorentz symmetry using the Faddeev--Popov procedure. We then study this model to investigate the impact of supersymmetry on the dynamical generation of $(3+1)$--dimensional spacetime.}

\FullConference{The 42nd International Symposium on Lattice Field Theory (LATTICE2025)\\
2-8 November 2025\\
Tata Institute of Fundamental Research, Mumbai, India\\}

\begin{document}
%\begin{flushright}
%KEK-TH-XXXX {\bf CAN IT BE HERE?}
%\end{flushright}
\maketitle
%----------------------------------------------------------------------------------------------------------------------------------
\section{Introduction}
%----------------------------------------------------------------------------------------------------------------------------------
Large--$N$ reduced matrix models have been proposed as a nonperturbative formulation of string theory. The type IIB matrix model, also known as the IKKT model \cite{hep-th/9612115}, is considered one of the most promising candidates. This model is formally defined through the dimensional reduction of the $(9+1)$--dimensional ${\cal N}=1$ super Yang--Mills theory to zero dimension. 
The spacetime does not exist a priori but rather emerges dynamically from the degrees of freedom of the matrices. Since superstring theory is defined in a $(9+1)$--dimensional spacetime, a central problem is to elucidate the mechanism governing the dynamical generation of our observed $(3+1)$--dimensional spacetime. This effect  has been conjectured to occur via the spontaneous symmetry breakdown (SSB) of the spatial SO($9$) symmetry.
The Lorentzian version of the type IIB matrix model encounters a significant sign problem, which originates from the complex phase\footnote{The Pfaffian $\text{Pf} {\cal M}'(A)$, resulting from the integration of the fermions in Eq.~\eqref{ferm_mass2}, is real. Furthermore, it remains positive for the dominant configurations contributing to the path integral. Consequently, its contribution to the sign problem is small compared to  $e^{iS}$.}  $e^{iS}$ in the partition function, where $S$ is the action of the model. Early investigations into the Lorentzian model used an approximation to circumvent this sign problem \cite{1108_1540}. Although this approximation yielded an expanding $(3+1)$--dimensional spacetime, the expanding dimensions of the space were found to be singular, described essentially by the Pauli matrices \cite{1904_05914}. 
This motivated several subsequent efforts  \cite{1904_05919,2210_17537} to simulate the Lorentzian version using the Complex Langevin Method (CLM)  \cite{Parisi:1983mgm,Klauder:1983sp}, which is a successful method to handle complex--action systems. Prior studies, reported in Ref.~\cite{2210_17537} and the references therein, reported the emergence of a $(1+1)$--dimensional spacetime; however, this result was later identified as an artifact stemming from large Lorentz boosts due to the Lorentz symmetry of the model.
In Refs.~\cite{2407_03491,2604_19836}, simulations were carried out by selecting an appropriate Lorentz frame. This method successfully eliminated the Lorentz boost artifact and provided evidence suggesting the emergence of $(3+1)$--dimensional spacetime.
In this paper we describe how to ``gauge--fix'' the Lorentz symmetry using the Faddeev--Popov procedure, following Ref.~\cite{2404_14045}. We perform numerical simulations of the type IIB matrix model by applying this ``gauge--fixing'' of the Lorentz symmetry in conjunction with the CLM. Our primary objective is to determine the mechanism by which the $(3+1)$--dimensional spacetime emerges dynamically from the type IIB matrix model.
The remainder of this paper is organized as follows. Section \ref{Sec_2} provides the definition of the Lorentzian version of the type IIB matrix model. Section \ref{Sec_3} presents the prescription for fixing the Lorentz symmetry via the Faddeev--Popov procedure. Section \ref{Sec_4} discusses the application of the Complex Langevin Method (CLM) to the type IIB matrix model. Numerical results obtained from the CLM simulation are presented in Section \ref{Sec_5}. Section \ref{Sec_6} is dedicated to a summary and discussion.

%----------------------------------------------------------------------------------------------------------------------------------
\section{Definition of the Lorentzian type IIB matrix model} \label{Sec_2}
%----------------------------------------------------------------------------------------------------------------------------------
The action of the type IIB matrix model \cite{hep-th/9612115} is given by $S=S_{\textrm{b}} + S_{\textrm{f}}$, where
\begin{eqnarray}
 S_{\textrm{b}} &=& - \frac{N}{4} \textrm{tr} ([A_{\mu}, A_{\nu}] [A^{\mu}, A^{\nu}] )\ , \label{IKKT_boson} \\
 S_{\textrm{f}} &=& - \frac{N}{2} \textrm{tr} \left( \psi_\alpha
 ({\cal C} \Gamma^{\mu})_{\alpha\beta}
 [A_{\mu}, \psi_\beta] \right) \ , \label{IKKT_fermion}
\end{eqnarray}
Here, $A_{\mu}$ ($\mu=0,1,\cdots,9$) and $\psi$ are $N \times N$ traceless Hermitian matrices. $\Gamma^{\mu}$, ${\cal C}$ ($\alpha, \beta=1,2,\cdots,16$) are the ten--dimensional gamma matrices and the charge conjugation matrix, respectively, which are obtained after the Weyl projection. We study the Lorentzian version throughout this paper, in which the indices $\mu, \nu$ are contracted with the Minkowski metric $\eta^{\mu \nu} = (-1,1,1,\cdots,1)$. The action has manifest SO($9,1$) symmetry, where $A_{\mu}$ and $\psi_{\alpha}$ transform as a vector and a Majorana--Weyl spinor, respectively.

Integrating out the fermionic matrices $\psi$, the partition function becomes
\begin{equation}
Z=\int dA d \psi\,  e^{i(S_\mathrm{b}+ S_\mathrm{f})} = \int dA\, e^{i S_\mathrm{b}}\, \mathrm{Pf} \mathcal{M}(A_0,A_i) \, . \label{IKKT_partition}
\end{equation}
The matrix ${\cal M}$ is a $16 (N^2-1) \times 16(N^2-1)$ anti--symmetric matrix, representing the linear transformation 
\begin{eqnarray}
 \psi_{\alpha} \to ({\cal M} \psi)_{\alpha} =  (\mathcal{C}\Gamma^{\mu})_{\alpha \beta} [A_{\mu}, \psi_{\beta}], \label{calM_def}
\end{eqnarray}
which acts on the linear space of traceless complex $N \times N$ matrices $\psi_{\alpha}$. 

In the type IIB matrix model, neither time nor space exists {\it a priori}. The eigenvalues of the bosonic matrices $A_{\mu}$ are interpreted as the spacetime coordinates, due to the ${\cal N}=2$ supersymmetry of the model.

The Pfaffian of the model $\textrm{Pf } {\cal M}$ favors lower dimensional configurations \cite{hep-th/0003223}. However, it vanishes when  the $A_{\mu}$ are nonzero only for two directions, thereby disfavoring $(1+1)$--dimensional or $(2+1)$--dimensional spacetime. (Note that the spatial matrices typically become exponentially larger than the temporal matrix $A_0$, which can therefore be assumed to be zero in this discussion.) We therefore expect that the effect of supersymmetry will play a crucial role in the dynamical generation of $(3+1)$--dimensional spacetime.

We define time by diagonalizing $A_0$ using an SU($N$) gauge transformation, such that the eigenvalues of  $A_0$ are arranged in ascending order:
\begin{eqnarray}
A_0=\text{diag}(\alpha_1,\alpha_2,\dots,\alpha_N),\quad \alpha_1\le\alpha_2\le\dots\le\alpha_N. \label{diagonal_gauge_A0}
\end{eqnarray}
In this SU($N$) basis, where $A_0$ is diagonalized as in Eq.~\eqref{diagonal_gauge_A0}, the spatial matrices $A_i$ exhibit a band--diagonal structure for typical configurations with expanding behaviors. Specifically, for a certain integer $n$, the $|(A_i)_{ab}|$ for $|a-b|>n$ are significantly smaller than the $|(A_i)_{ab}|$ for $|a-b| \leq n$ \cite{1108_1540}. This observation motivates the introduction of $n \times n$ matrices $\bar{A}_i(t)$ defined as $(\bar{A}_i)_{pq}(t_a) = (A_i)_{a+p, a+q}$, with $p,q=1,2,\dots,n$ and $a=0,1,\dots,N-n$. The time $t_{a}$ is then defined as:
\begin{equation}
t_0=0, \ t_a = \sum_{b=1}^{a} | \bar{\alpha}_{b+1} -  \bar{\alpha}_b |, \textrm{ where } \bar{\alpha}_{b+1} = \frac{1}{n} \sum_{q=1}^{n} \alpha_{b+q}\, . 
\label{time_def}
\end{equation}

As an order parameter for the  SSB of the SO($9$) symmetry, we define the $9 \times 9$ symmetric tensor
\begin{eqnarray}
 T_{ij}(t) = \frac{1}{n} \textrm{tr} ({\bar A}_i (t) {\bar A}_j (t)). \label{Tij_tensor}
\end{eqnarray}
We denote the eigenvalues of $T_{ij} (t)$ as $\lambda_{T,k} (t)$ $(k=1,2,\cdots,9)$ ordered in descending magnitude  $\lambda_{T,1} (t) >  \lambda_{T,2} (t) > \dots >  \lambda_{T,9} (t)$. If $ \langle \lambda_{T,1} (t) \rangle, \ \dots,   \langle \lambda_{T,d} (t) \rangle$  are larger  than the remaining eigenvalues, this signals spontaneous symmetry breaking from SO($9$) to SO($d$), which implies the dynamical generation of a $d$--dimensional space.

%----------------------------------------------------------------------------------------------------------------------------------
\section{``Gauge--fixing'' of the Lorentz symmetry} \label{Sec_3}
%----------------------------------------------------------------------------------------------------------------------------------
It turns out that diagonalizing $A_0$ as in Eq.~\eqref{diagonal_gauge_A0} using the SU($N$) gauge transformation is insufficient to fully fix the Lorentz symmetry. A method to fix the Lorentz symmetry via the Faddeev--Popov procedure has been proposed in Ref.~\cite{2404_14045}. In the following discussion, we refer to this prescription as the ``gauge--fixing'' of the Lorentz symmetry.

The gauge--fixing condition is derived by choosing a frame that minimizes the quantity ${\cal T}$ with respect to Lorentz transformations on each sampled configuration \cite{2404_14045,2407_03491} as
\begin{eqnarray}
{\cal Q} = {\cal T}-{\cal U}\, ,\quad \textrm{ where }\quad {\cal T} = \textrm{tr} (A_0^2)\, ,\quad {\cal U} = \sum_{i=1}^9 \textrm{tr} (A_i^2), \label{Lorenrtz_invariant_Q}
\end{eqnarray}
where ${\cal Q}$ is Lorentz invariant, and ${\cal T}$ and ${\cal U}$ are varied using the Lorentz boosts:
\begin{eqnarray}
 \left( \begin{array}{c} A_0' \\ A_i' \end{array} \right) =  \left( \begin{array}{cc} \cosh \sigma & \sinh \sigma \\ \sinh \sigma & \cosh \sigma  \end{array} \right)  \left( \begin{array}{c} A_0 \\ A_i \end{array} \right) \, .\label{Lorentz_tr1}
\end{eqnarray}

Minimizing  ${\cal T}$ leads to the condition $\textrm{tr} (A_0 A_i) = 0$, for all $i=1,\ldots,9$. This condition is implemented nonperturbatively through the Faddeev--Popov procedure, yielding the partition function:
\begin{eqnarray}
 Z_{\textrm{g.f.}}= \int dA\, e^{i S_\textrm{b}}\, \mathrm{Pf} \mathcal{M}\, \,\Delta_{\textrm{FP}} \,\prod_{i=1}^9 \delta(\textrm{tr} (A_0 A_i)) \ .  \label{GF2}
\end{eqnarray}

The Faddeev--Popov determinant $\Delta_{\textrm{FP}}$ is defined as $\Delta_{\textrm{FP}}=\det \Omega$, where  $\Omega$ is a  $9 \times 9$ matrix with components:
\begin{eqnarray} 
\Omega_{ij} = \textrm{tr} (A_0^2) \delta_{ij} + \textrm{tr} (A_i A_j), \quad (i,j=1,2,\dots,9). \label{GF3} 
\end{eqnarray}

%----------------------------------------------------------------------------------------------------------------------------------
\section{Complex Langevin method} \label{Sec_4}
%----------------------------------------------------------------------------------------------------------------------------------
The CLM \cite{Parisi:1983mgm,Klauder:1983sp} is a promising  numerical technique for simulating systems with complex actions. It operates by formulating stochastic differential equations for complexified variables, which can then be utilized to calculate expectation values under specific convergence conditions.

Consider a general model defined by the partition function  $Z=\int_M dx\, w(x)$, where $x$ represents the real degrees of freedom,  $M$  is the manifold of these coordinates, and  $w(x)$ is a complex--valued weight function. In the CLM, the real variables are complexified,  $x \to z$, where $z$ resides in the complexification  $M^c$ of $M$. One then solves the complex Langevin equation with respect to the fictitious Langevin time  $\sigma$:
\begin{eqnarray}
\label{Langevin_eq}
\frac{d z_k}{d \sigma}= \frac{1}{w(z)} \frac{\partial w(z)}{\partial z_k} +{\eta_k(\sigma)}.
\end{eqnarray}

The first term on the right--hand side of Eq.~\eqref{Langevin_eq} is the ``drift term,'' while the second term represents real Gaussian noise with a probability distribution proportional to  $\exp \left( - \frac{1}{4} \int d\sigma \sum_k [\eta_k(\sigma)]^2 \right)$. 

The solution to the complex Langevin equation is given by the probability distribution $P(z;\sigma)$, which satisfies a corresponding Fokker--Planck equation. Under certain conditions, this distribution converges to a stationary distribution  $P[z]$ as $\sigma\to\infty$. If  ${\cal O}[x]$  is an observable, and if the condition   $\int_{M^c} dz\, {\cal O}[z] P[z] = \int_M dx\, {\cal O}[x] w[x]$  is satisfied, then the expectation value of   ${\cal O}[x]$ can be evaluated as the average over the Langevin trajectory:
%% %%%%%%%%%%%%%%%
\begin{eqnarray}
\langle {\cal O}[x] \rangle = \frac{1}{\sigma} \int_{\sigma_0}^{\sigma_0+\sigma} d\sigma' {\cal O}[x(\sigma')]\, ,\label{kna001}
\end{eqnarray}
%% %%%%%%%%%%%%%%%
where  $\sigma_0$ is the time at which the evolution has reached equilibrium. The holomorphy of both the weight function  $w(z)$ and the observable  ${\cal O}[z]$ is essential for establishing the validity of Eq.~\eqref{kna001}. 

In Ref.~\cite{1904_05919}, a method was proposed to incorporate the time order defined by Eq.~\eqref{diagonal_gauge_A0} for the complexified eigenvalues  $\alpha_i$. This approach involves introducing auxiliary variables  $\tau_i$ (where $i=1,2,\cdots,N-1$) such that the eigenvalues are parameterized as:
\begin{eqnarray}
 \alpha_1 = 0, \ \alpha_k = \sum_{i=1}^{k-1} e^{\tau_i}. \label{time_order}
\end{eqnarray}
Using the shift symmetry $A_0 \to A_0 + \textrm{(const.)} {\bf 1}_N$, we apply the shift $A_0 \to A_0 - \left( \frac{1}{N} \textrm{tr} A_0 \right) {\bf 1}_N$ to ensure  $A_0$ is traceless. This symmetry allows us to present plots where the time eigenvalues  $\alpha_a$ and the time $t$ are centered at $0$. Together with the diagonalization of $A_0$, it introduces the following term into the effective action:
\begin{eqnarray}
 S_{A\textrm{diag}} = - \log \prod_{1 \leq a<b \leq N} (\alpha_a - \alpha_b)^2 - \sum_{a=1}^{N-1} \tau_a. \label{time_order2}
\end{eqnarray}

The near--zero modes of the Dirac operator defined in Eq.~\eqref{IKKT_fermion} lead to the singular--drift problem \cite{1504_08359,1606_07627,0912_3360}, which is known to make the complex Langevin simulation unreliable. To mitigate this issue, we introduce a fermionic mass term:
\begin{eqnarray}
 S_{m_\textrm{f}} = \frac{-N}{2} i m_{\textrm{f}} \textrm{tr} {\bar \psi} (\Gamma^7 (\Gamma^8)^\dag \Gamma^9) \psi \, . \label{ferm_mass}
\end{eqnarray}
This mass term breaks the  SO($9$) symmetry down to $\textrm{SO}(6) \times \textrm{SO}(3)$. For a real parameter $m_{\textrm{f}}$, this term effectively shifts the eigenvalues of the matrix  ${\cal M}$ away from the origin. Consequently, the fermionic degrees of freedom are integrated out to yield the modified Pfaffian $\textrm{Pf } {\cal M}'$:
\begin{eqnarray}
 \int d \psi e^{i (S_{\textrm{f}} + S_{m_\textrm{f}})} = \textrm{Pf } {\cal M}'. \label{ferm_mass2}
\end{eqnarray}

In the limit  $m_{\textrm{f}} \to \infty$, the fermionic degrees of freedom decouple, and the system reduces to the pure bosonic model. Introducing a non--zero $m_{\textrm{f}}$ weakens the effect of supersymmetry. To compensate for this weakening and to control the bosonic quantum fluctuations in a manner that mimics supersymmetry, we introduce the following bosonic mass term:
\begin{eqnarray}
 S_{\gamma} = \frac{N\gamma}{2} \textrm{tr} \left\{ (A_0)^2 - \sum_{i=1}^{{\tilde d}} (A_i)^2 - \xi \sum_{i={\tilde d}+1}^9 (A_i)^2 \right\}. \label{xi_massterm}
\end{eqnarray}
Here, ${\tilde d}$ is an integer  less than or equal to $6$, since the fermionic mass term \eqref{ferm_mass} already reduces the SO($9$) symmetry to $\textrm{SO}(6) \times \textrm{SO}(3)$. Consequently, the combination of $S_{m_\textrm{f}}$ \eqref{ferm_mass} and $S_{\gamma}$  \eqref{xi_massterm} breaks the SO($9$) spatial symmetry down to $\textrm{SO}({\tilde d}) \times \textrm{SO}(3)$. In the subsequent analysis, we set $\tilde d=6$ to focus on exploring the remaining spatial symmetry.

At $\xi=1$, this term reduces to the Lorentz invariant mass term  $\frac{-N\gamma}{2} \textrm{tr} (A^{\mu} A_{\mu})$ \cite{1911_08132}. Furthermore, when combined with the Myers term, the model becomes the supersymmetric deformation of the type IIB matrix model developed in Ref.~\cite{hep-th/0205213}, provided the parameters are set as $m_{\textrm{f}} = \frac{\mu}{4}$, $\gamma = \frac{-\mu^2}{32}$, $\xi=3$, and ${\tilde d}=6$, with $\mu$ being a real parameter.

In total, the effective action  $S_{\textrm{eff}}$ used for the numerical simulation is defined by the partition function:
 \begin{eqnarray}
 Z_{\textrm{eff}}= \int dA_i d \tau_k e^{-S_{\textrm{eff}}} = \int dA_i d \tau_k\, e^{i (S_\textrm{b} +  S_{\gamma}) - S_{A\textrm{diag}}}\,\, \mathrm{Pf} {\cal M}' \,\,\Delta_{\textrm{FP}} \,\,\prod_{i=1}^9 \delta(\textrm{tr} (A_0 A_i)) \ .  \label{Seff_total}
\end{eqnarray}
Here, the terms $S_{\textrm{b}}$, $S_{\gamma}$, $S_{A\textrm{diag}}$, ${\cal M}'$, and $\Delta_{\textrm{FP}}$ are defined by Eqs.~\eqref{IKKT_boson},  \eqref{xi_massterm}, \eqref{time_order2},  \eqref{ferm_mass2} and \eqref{GF3}, respectively. We are ultimately interested in taking the limits  $m_{\textrm{f}}\to 0$ and $\gamma \to 0$ after the $N\to \infty$ limit. 

To implement the CLM, we complexify $\tau_a$ and allow the $A_i$ to take values in  SL($N,\mathbb{C}$).  The corresponding set of complex Langevin equations is given by: 
\begin{equation}
\frac{d(A_i)_{ab}}{d\sigma}
=- \ {\frac{\partial S_\mathrm{eff}}{\partial (A_i)_{ba}}} +{(\eta_i)_{ab}(\sigma)}\, , \quad 
\frac{d\tau_a}{d\sigma}
=- \ {\frac{\partial S_\mathrm{eff}}{\partial \tau_a}} +{\eta_a(\sigma)}\, . 
\label{CL_eq_LIKKT}
\end{equation}
Here,  $(\eta_i)_{ab} (\sigma)$ and $\eta_a (\sigma)$ represent the noise terms. Specifically, $\eta_i (\sigma)$ are Hermitian matrices, and  $\eta (\sigma)$ are real vectors, with the respective probability distributions proportional to \\ \noindent $\exp \left( - \frac{1}{4} \int d \sigma \textrm{tr} \eta_i^2 (\sigma) \right)$ and $\exp \left( - \frac{1}{4} \int d \sigma \eta_a^2 (\sigma) \right)$. 

%----------------------------------------------------------------------------------------------------------------------------------
\section{Results} \label{Sec_5}
%----------------------------------------------------------------------------------------------------------------------------------
%%\begin{eqnarray}
%% N=32, \ n=8, \ m_{\textrm{f}} = 2, \ \gamma = 6, \ \xi=10, \ {\tilde d}=6. \label{our_parameters}
%%\end{eqnarray}
\begin{figure}[htbp]
\vspace*{-2mm}
\centering % \begin{center}/\end{center} takes some additional vertical space
%\includegraphics[width=0.490\hsize]{tsolN032gm06-00alp00100xi10-00dt2ms00-00ph01-00_Re_2dinitial.eps} %TL(上段左) \includegraphics[width=0.490\hsize]{tsolN032gm06-00alp00100xi10-00dt6ms02-00ph01-00_Re_2d.eps} %TR(上段右)
%\includegraphics[width=0.490\hsize]{tsolN032gm06-00alp00100xi10-00dt3ms00-00ph01-00_Re_3dinitial.eps} %ML(中段左) \includegraphics[width=0.490\hsize]{tsolN032gm06-00alp00100xi10-00dt6ms02-00ph01-00_Re_3d.eps} %MR(中段右)
%\includegraphics[width=0.490\hsize]{tsolN032gm06-00alp00100xi10-00dt4ms00-00ph01-00_Re_4dinitial.eps} %BL(下段左) \includegraphics[width=0.490\hsize]{tsolN032gm06-00alp00100xi10-00dt6ms02-00ph01-00_Re_4d.eps} %BR(下段右)
% 横方向のすき間調整（好みで変更）
\setlength{\tabcolsep}{0mm}

\begin{tabular}{ccc}
  % 上段: TL -> TR
  \raisebox{-0.5\height}{%
    \includegraphics[width=0.470\hsize]{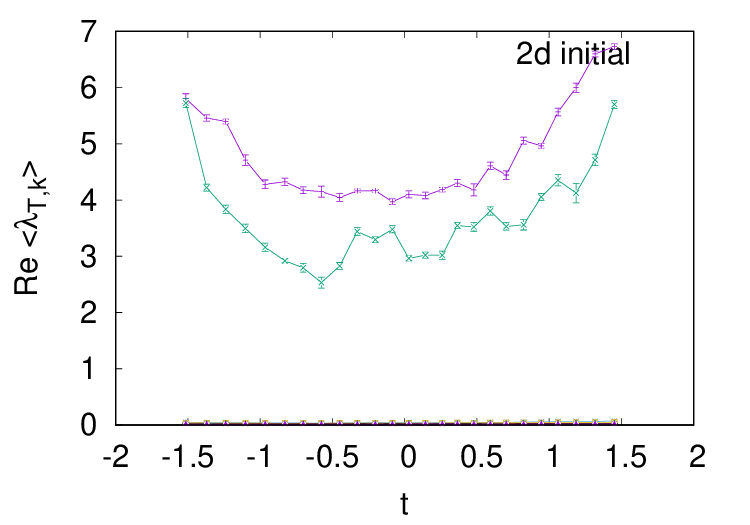}} %TL
  &
  \raisebox{-0.5\height}{\Huge$\Rightarrow$}
  &
  \raisebox{-0.5\height}{%
    \includegraphics[width=0.470\hsize]{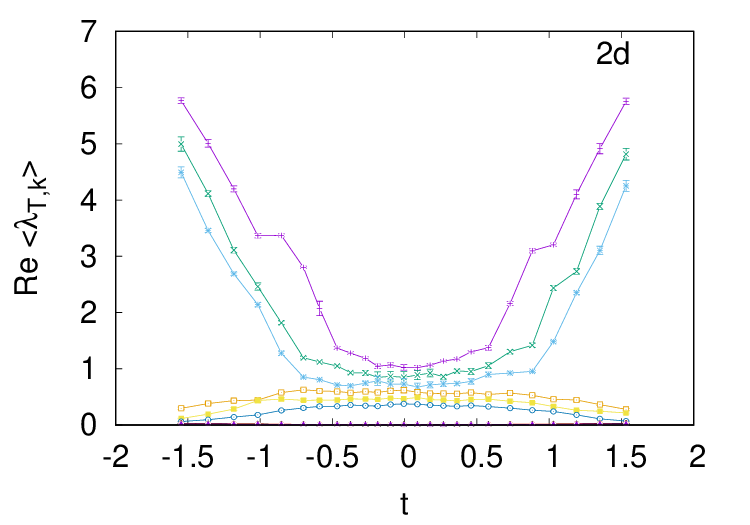}} \\[2mm] %TR

  % 中段: ML -> MR
  \raisebox{-0.5\height}{%
    \includegraphics[width=0.470\hsize]{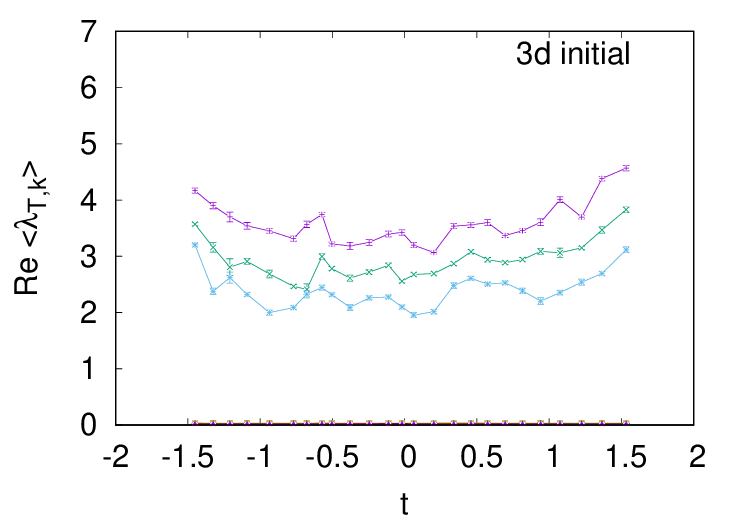}} %ML
  &
  \raisebox{-0.5\height}{\Huge$\Rightarrow$}
  &
  \raisebox{-0.5\height}{%
    \includegraphics[width=0.470\hsize]{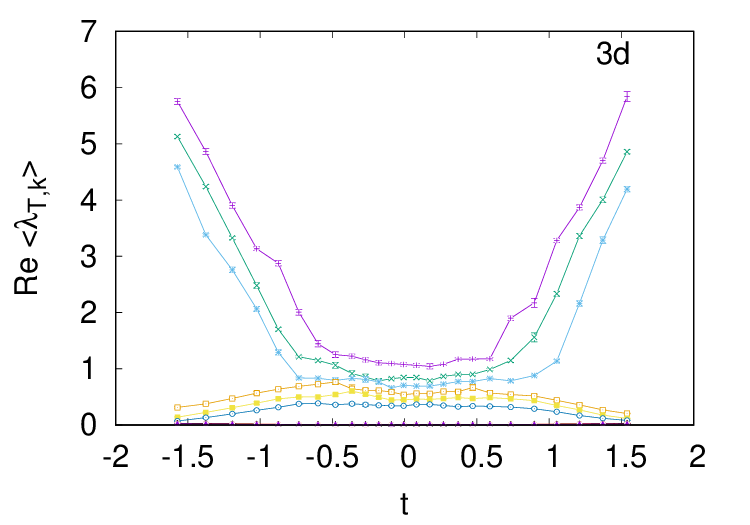}} \\[2mm] %MR

  % 下段: BL -> BR
  \raisebox{-0.5\height}{%
    \includegraphics[width=0.470\hsize]{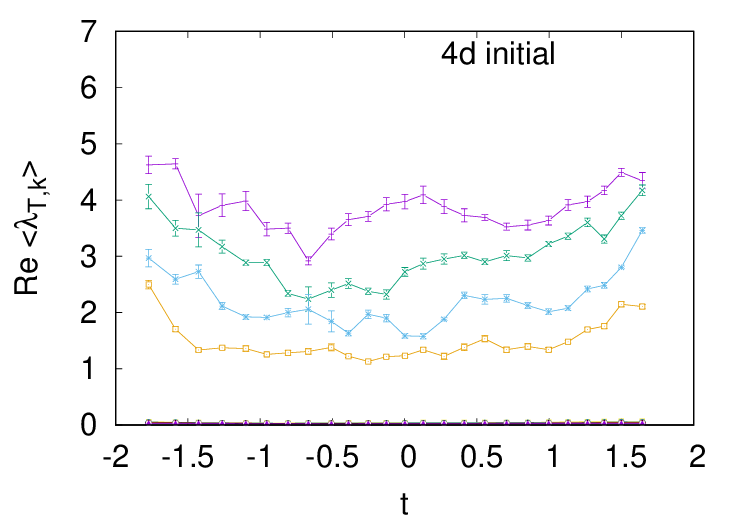}} %BL
  &
  \raisebox{-0.5\height}{\Huge$\Rightarrow$}
  &
  \raisebox{-0.5\height}{%
    \includegraphics[width=0.470\hsize]{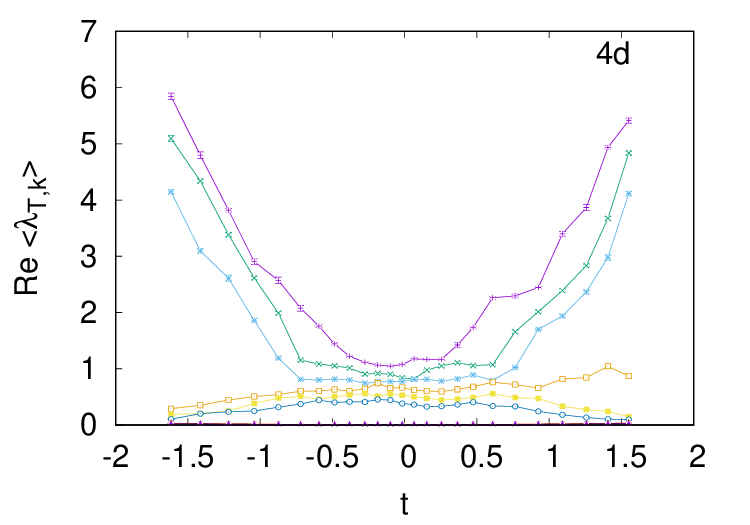}} %BR
\end{tabular}

    \caption{The real parts of the expectation values of the eigenvalues $\textrm{Re} \langle \lambda_{T,k} (t) \rangle$ ($k=1,2,\cdots,9$) are plotted  as a function of the time $t$ for simulations initialized with configurations corresponding to $2$--dimensional (Top row), $3$--dimensional (Middle row), and $4$--dimensional (Bottom row) spaces. The left panels display  $\textrm{Re} \langle \lambda_{T,k} (k) \rangle$ for the initial configurations, while the right panels show the results for the thermalized configurations.}
    \label{Tij_expansion}
    \vspace*{-2mm}
\end{figure}

In this section, we present the results of the CLM. The parameter set used for the following simulations is $N=32$, $n=8$, $m_{\textrm{f}} = 2$, $\gamma = 6$, $\xi=10$, and ${\tilde d}=6$. We use initial configurations that represent $2$--, $3$--, and $4$--dimensional spaces. These configurations were generated by solving the Langevin equations \eqref{CL_eq_LIKKT} for the ``bosonic model'' (where $\textrm{Pf } {\cal M}$ is omitted) with ${\tilde d}=2,3,4$ respectively, and $N=32$, $n=8$, $\gamma = 6$, $\xi=10$. Since  $A_i\in $ SL($N,\mathbb{C}$), the eigenvalues $\lambda_{T,i} (t)$ of  $ T_{ij}(t)$ defined in Eq.~\eqref{Tij_tensor} are not real. These eigenvalues are ordered according to their real parts, and subsequently, their expectation values  $\langle \lambda_{T,k} (t) \rangle$  ($k=1,2,\cdots,9$) are calculated. $\textrm{Re} \langle \lambda_{T,k} (t) \rangle$ ($k=1,2,\cdots,9$), for these starting points are shown in Figure \ref{Tij_expansion} (Left). 

Figure~\ref{Tij_expansion} (Right) presents the real parts of the expectation values,  $\textrm{Re} \langle \lambda_{T,k} (t) \rangle$ ($k=1,2,\cdots,9$), plotted against time $t$ for the thermalized configurations obtained from the $2$--, $3$--, and $4$--dimensional initial configurations. A notable finding is that, irrespective of the dimensionality of the initial configurations, the thermalized results consistently signal the emergence of an expanding $3$--dimensional space. Specifically, at early times, all nine $\textrm{Re} \langle \lambda_{T,k} (t) \rangle$ are small, which corresponds to a small $9$--dimensional space.  Subsequently, {\it three} of these eigenvalues begin to grow at late times, which signals the SSB of the  SO($9$) symmetry down to SO($3$). This observed behavior is non--trivial; starting from a random configuration, for instance, typically yields no SSB, resulting instead in a $6$--dimensional expansion. The fact that commencing from initial configurations of varying dimensionality leads to the same thermalized $3$--dimensional expansion provides strong confidence in the ergodicity of our Complex Langevin simulations.

%------------------------------------------------ XXXX -----------------------------------------------------------------------
We define the extent of space $R^2 (t)$ and its phase $\theta_s (t)$ at time $t$ as:
\begin{eqnarray}
 R^2 (t) = \left\langle \frac{1}{n} \textrm{tr} \sum_{i=1}^{9} ({\bar A}_i(t))^2 \right\rangle  = e^{2i \theta_s (t)} |R^2 (t)|\, . \label{R_sq_def}
\end{eqnarray}
A phase of $\theta_\mathrm{s}(t) \sim 0$ suggests the emergence of real space in the Lorentzian model, while $\theta_\mathrm{s}(t) \sim \frac{\pi}{8}$ indicates Euclidean space.
%------------------------------------------------ XXXX -----------------------------------------------------------------------

\begin{figure}[htbp]
\vspace*{-4mm}
\centering % \begin{center}/\end{center} takes some additional vertical space
    \includegraphics[width=0.490\hsize]{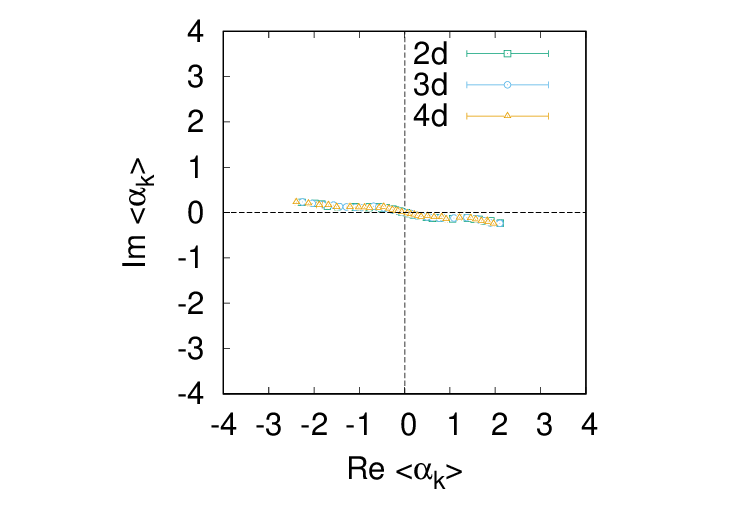}
    \includegraphics[width=0.490\hsize]{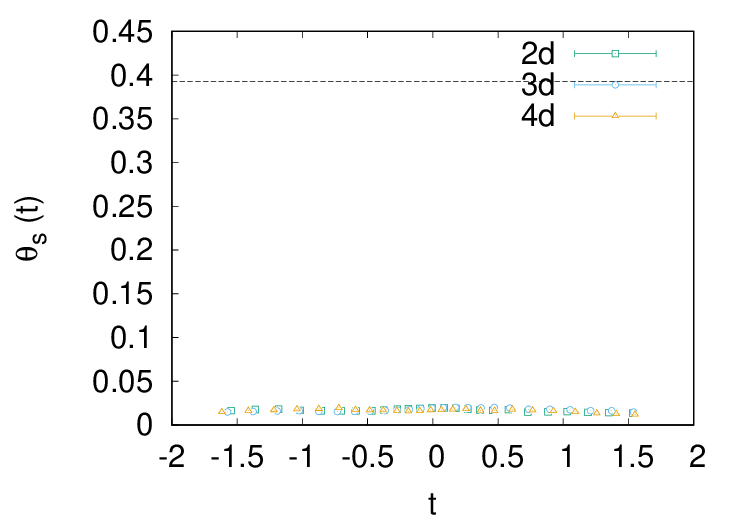}
     \caption{(Left)  The expectation values  $\left\langle \alpha_a \right\rangle$ are plotted in the complex plane.  (Right) The phase of space, $\theta_s (t)$, is plotted as a function of time $t$. The dotted line indicates the value $\frac{\pi}{8}$, which would correspond to the emergence of Euclidean space instead of real space. The labels  2d, 3d, and 4d correspond to the thermalized configurations obtained from the  $2$--, $3$--, and $4$--dimensional  initial configurations, respectively.}
    \label{spacetime_real}
    \vspace*{-4mm}
\end{figure}

The vacuum expectation values  $\left\langle \alpha_a \right\rangle$ are plotted in the complex plane in Figure \ref{spacetime_real} (Left). In all observed cases, these values lie close to the real axis, which strongly suggests the emergence of real time. Figure \ref{spacetime_real} (Right) displays the phase of space  $\theta_s (t)$, defined by Equation \eqref{R_sq_def}, as a function of time $t$. These phase values remain consistently much smaller than $\frac{\pi}{8}$, which implies that the emerging space is real, as opposed to Euclidean \cite{2112_15368}.

\begin{figure}[htbp]
\vspace*{-4mm}
\centering % \begin{center}/\end{center} takes some additional vertical space
    \includegraphics[width=0.490\hsize]{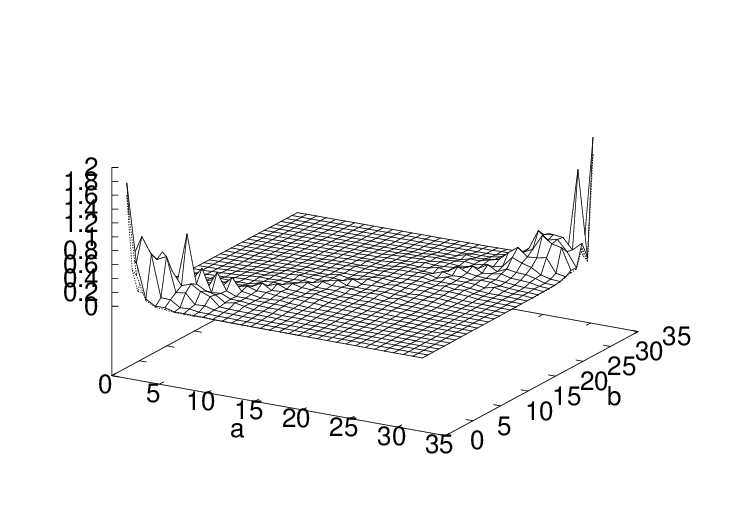}
\vspace*{-4mm}
    \caption{The magnitude ${\cal A}_{ab}$ is plotted for the thermalized configuration obtained starting from the $3$--dimensional initial configurations. The plot illustrates the band-diagonal structure of the spatial matrices, which is similarly observed for simulations originating from the $2$-- and $4$--dimensional initial configurations. }
    \label{band_diagonal_plot}
    \vspace{-4mm}
\end{figure}

To probe the band--diagonal structure of the spatial matrices $A_i$, we plot the magnitude  ${\cal A}_{ab}$,  defined by:
\begin{eqnarray}
 {\cal A}_{ab} = \frac{1}{9} \textrm{Re} \left\langle \sum_{i=1}^9 (A_i)_{ab} (A_i)_{ba} \right\rangle  \, , \quad (a,b=1,2,\dots,N)\, , \label{Apq_def}
\end{eqnarray}
in Figure \ref{band_diagonal_plot} for the simulation starting from the $3$--dimensional initial configurations. This visualization clearly suggests that the matrices  $A_i$ possess a band--diagonal structure. Specifically, the values of  ${\cal A}_{ab}$ for off--diagonal elements satisfying  $|a-b|>n$ are significantly smaller than those where  $|a-b| \leqq n$. A similar band--diagonal structure is also observed for simulations originating from the $2$--dimensional  and $4$--dimensional  initial configurations.

%\end{minipage}

%----------------------------------------------------------------------------------------------------------------------------------
\section{Summary and discussion} \label{Sec_6}
%----------------------------------------------------------------------------------------------------------------------------------
In this work, we have investigated the Lorentzian version of the type IIB matrix model, a proposed non-perturbative formulation of superstring theory, using the Complex Langevin Method (CLM) to overcome the sign problem.
To eliminate the artifacts associated with the Lorentz boost, we studied the model with the Lorentz symmetry ``gauge--fixed'' using the Faddeev--Popov procedure \cite{2404_14045}. Our results indicate that for the chosen fermionic mass parameter, $m_{\textrm{f}}=2$ (the coefficient of the mass term \eqref{ferm_mass}), the space maintains SO($9$) symmetry at early times  but spontaneously breaks down to SO($3$) at a critical time. This interesting effect was consistently observed across simulations starting from  $2$--, $3$--, and $4$--dimensional initial configurations. Furthermore, this result was achieved by selecting a sufficiently small value for $m_{\textrm{f}}$, which suggests that the dynamical generation of the $3$--dimensional space is directly attributable to the impact of supersymmetry.
There remain several future problems that warrant investigation. The work reported herein utilized a small matrix size, $N=32$. It is therefore essential to perform large--scale numerical simulations for larger $N$ to  fully explore the large--$N$ limit, followed by the limits  $\gamma \to 0$, $m_{\textrm{f}} \to 0$ \cite{inprogress}. Furthermore, it would be interesting to study the Lorentzian version of the type IIB matrix model incorporating a supersymmetry--preserving mass term \cite{hep-th/0205213}. The Euclidean counterpart of this model, featuring the supersymmetry--preserving mass term, has already been explored in previous works \cite{2209_10494,2308_03607,2507_18472}. 
Finally, the Lefschetz thimble method represents another promising alternative for addressing systems afflicted by the sign problem \cite{1001_2933,2501_17798}. While this method enables us to determine the dominant configurations more reliably than the CLM, it requires a lot of computational efforts to perform simulations at large $N$. This is also important as a future direction.

\vspace{-0.4em}
\section*{Acknowledgements}
A.\;T. was supported in part by Grant--in--Aid (Nos. 21K03532, 25K07319) from Japan Society for the Promotion of Science. This work was supported by MEXT as ``Program for Promoting Researches on the Supercomputer Fugaku'' (Simulation for basic science: approaching the new quantum era, PMXP1020230411) and JICFuS. Part of the computations were performed using supercomputer Fugaku provided by the RIKEN Center for Computational Science (Project ID: hp250224).
This work also used computational resources of Oak--bridge--CX provided by the University of Tokyo (Project IDs: hp200106, hp210094, hp220074, hp230149), Grand Chariot provided by Hokkaido University (Project IDs: hp230149, hp240117) and Flow provided by Nagoya University (Project ID hp250140) through the HPCI System Research Project.
Numerical computations were also carried out on PC clusters in KEK Computing Research Center.  
This work was supported by computational time granted by the Greek Research and Technology Network (GRNET) in the National HPC facility ARIS, under the project IDs LIIB2, QCMM, and QCMMII.

\vspace{-0.4em}
\setlength{\baselineskip}{10.5pt}
\bibliographystyle{JHEP2}
\bibliography{bib.bib}

@article{hep-th/9612115,
    author = "Ishibashi, N. and Kawai, H. and Kitazawa, Y. and Tsuchiya, A.",
    title = "{A Large N reduced model as superstring}",
    eprint = "hep-th/9612115",
    archivePrefix = "arXiv",
    reportNumber = "KEK-TH-503",
    doi = "10.1016/S0550-3213(97)00290-3",
    journal = "Nucl. Phys. B",
    volume = "498",
    pages = "467--491",
    year = "1997"
}

@article{1108_1540,
    author = "Kim, Sang-Woo and Nishimura, Jun and Tsuchiya, Asato",
    title = "{Expanding (3+1)-dimensional universe from a Lorentzian matrix model for superstring theory in (9+1)-dimensions}",
    eprint = "1108.1540",
    archivePrefix = "arXiv",
    primaryClass = "hep-th",
    reportNumber = "KEK-TH-1484, OU-HET-720-2011",
    doi = "10.1103/PhysRevLett.108.011601",
    journal = "Phys. Rev. Lett.",
    volume = "108",
    pages = "011601",
    year = "2012"
}

@article{1904_05914,
    author = "Aoki, Toshihiro and Hirasawa, Mitsuaki and Ito, Yuta and Nishimura, Jun and Tsuchiya, Asato",
    title = "{On the structure of the emergent 3d expanding space in the Lorentzian type IIB matrix model}",
    eprint = "1904.05914",
    archivePrefix = "arXiv",
    primaryClass = "hep-th",
    reportNumber = "KEK-TH-2110",
    doi = "10.1093/ptep/ptz092",
    journal = "PTEP",
    volume = "2019",
    number = "9",
    pages = "093B03",
    year = "2019"
}

@article{1904_05919,
    author = "Nishimura, Jun and Tsuchiya, Asato",
    title = "{Complex Langevin analysis of the space-time structure in the Lorentzian type IIB matrix model}",
    eprint = "1904.05919",
    archivePrefix = "arXiv",
    primaryClass = "hep-th",
    reportNumber = "KEK-TH-2119",
    doi = "10.1007/JHEP06(2019)077",
    journal = "JHEP",
    volume = "06",
    pages = "077",
    year = "2019"
}

@article{2210_17537,
    author = "Anagnostopoulos, Konstantinos N. and Azuma, Takehiro and Hatakeyama, Kohta and Hirasawa, Mitsuaki and Ito, Yuta and Nishimura, Jun and Papadoudis, Stratos Kovalkov and Tsuchiya, Asato",
    title = "{Progress in the numerical studies of the type IIB matrix model}",
    eprint = "2210.17537",
    archivePrefix = "arXiv",
    primaryClass = "hep-th",
    reportNumber = "KEK-TH-2470",
    doi = "10.1140/epjs/s11734-023-00849-x",
    journal = "Eur. Phys. J. ST",
    volume = "232",
    number = "23-24",
    pages = "3681--3695",
    year = "2023"
}

@article{Parisi:1983mgm,
    author = "Parisi, G.",
    title = "{On complex probabilities}",
    doi = "10.1016/0370-2693(83)90525-7",
    journal = "Phys. Lett. B",
    volume = "131",
    pages = "393--395",
    year = "1983"
}

@article{Klauder:1983sp,
    author = "Klauder, John R.",
    title = "{Coherent State Langevin Equations for Canonical Quantum Systems With Applications to the Quantized Hall Effect}",
    reportNumber = "Print-83-0902 (BTL)",
    doi = "10.1103/PhysRevA.29.2036",
    journal = "Phys. Rev. A",
    volume = "29",
    pages = "2036--2047",
    year = "1984"
}

@article{hep-th/0003223,
    author = "Nishimura, Jun and Vernizzi, Graziano",
    title = "{Spontaneous breakdown of Lorentz invariance in IIB matrix model}",
    eprint = "hep-th/0003223",
    archivePrefix = "arXiv",
    reportNumber = "NBI-HE-00-15",
    doi = "10.1088/1126-6708/2000/04/015",
    journal = "JHEP",
    volume = "04",
    pages = "015",
    year = "2000"
}

@article{2404_14045,
    author = "Asano, Yuhma and Nishimura, Jun and Piensuk, Worapat and Yamamori, Naoyuki",
    title = "{Defining the Type IIB Matrix Model without Breaking Lorentz Symmetry}",
    eprint = "2404.14045",
    archivePrefix = "arXiv",
    primaryClass = "hep-th",
    reportNumber = "UTHEP-787, KEK-TH-2617",
    doi = "10.1103/PhysRevLett.134.041603",
    journal = "Phys. Rev. Lett.",
    volume = "134",
    number = "4",
    pages = "041603",
    year = "2025"
}

@article{2407_03491,
    author = "Hirasawa, Mitsuaki and Anagnostopoulos, Konstantinos N. and Azuma, Takehiro and Hatakeyama, Kohta and Nishimura, Jun and Papadoudis, Stratos and Tsuchiya, Asato",
    title = "{The effects of SUSY on the emergent spacetime in the Lorentzian type IIB matrix model}",
    eprint = "2407.03491",
    archivePrefix = "arXiv",
    primaryClass = "hep-th",
    reportNumber = "KEK-TH-2636, KUNS-3007",
    doi = "10.22323/1.463.0257",
    journal = "PoS",
    volume = "CORFU2023",
    pages = "257",
    year = "2024"
}

@article{1504_08359,
    author = "Nishimura, Jun and Shimasaki, Shinji",
    title = "{New Insights into the Problem with a Singular Drift Term in the Complex Langevin Method}",
    eprint = "1504.08359",
    archivePrefix = "arXiv",
    primaryClass = "hep-lat",
    reportNumber = "KEK-TH-1816",
    doi = "10.1103/PhysRevD.92.011501",
    journal = "Phys. Rev. D",
    volume = "92",
    number = "1",
    pages = "011501",
    year = "2015"
}

@article{1606_07627,
    author = "Nagata, Keitaro and Nishimura, Jun and Shimasaki, Shinji",
    title = "{Argument for justification of the complex Langevin method and the condition for correct convergence}",
    eprint = "1606.07627",
    archivePrefix = "arXiv",
    primaryClass = "hep-lat",
    reportNumber = "KEK-TH-1911",
    doi = "10.1103/PhysRevD.94.114515",
    journal = "Phys. Rev. D",
    volume = "94",
    number = "11",
    pages = "114515",
    year = "2016"
}

@article{1911_08132,
    author = "Hatakeyama, Kohta and Matsumoto, Akira and Nishimura, Jun and Tsuchiya, Asato and Yosprakob, Atis",
    title = "{The emergence of expanding space{\textendash}time and intersecting D-branes from classical solutions in the Lorentzian type IIB matrix model}",
    eprint = "1911.08132",
    archivePrefix = "arXiv",
    primaryClass = "hep-th",
    reportNumber = "KEK-TH-2169",
    doi = "10.1093/ptep/ptaa042",
    journal = "PTEP",
    volume = "2020",
    number = "4",
    pages = "043B10",
    year = "2020"
}

@article{hep-th/0205213,
    author = "Bonelli, Giulio",
    title = "{Matrix strings in pp wave backgrounds from deformed superYang-Mills theory}",
    eprint = "hep-th/0205213",
    archivePrefix = "arXiv",
    reportNumber = "ULB-TH-02-15",
    doi = "10.1088/1126-6708/2002/08/022",
    journal = "JHEP",
    volume = "08",
    pages = "022",
    year = "2002"
}

@article{2112_15368,
    author = "Hatakeyama, Kohta and Anagnostopoulos, Konstantinos and Azuma, Takehiro and Hirasawa, Mitsuaki and Ito, Yuta and Nishimura, Jun and Papadoudis, Stratos and Tsuchiya, Asato",
    title = "{Relationship between the Euclidean and Lorentzian versions of the type IIB matrix model}",
    eprint = "2112.15368",
    archivePrefix = "arXiv",
    primaryClass = "hep-lat",
    reportNumber = "KEK-TH-2373",
    doi = "10.22323/1.396.0341",
    journal = "PoS",
    volume = "LATTICE2021",
    pages = "341",
    year = "2022"
}

@misc{inprogress,
    author = "Anagnostopoulos, Konstantinos N. and Azuma, Takehiro and Hirasawa, Mitsuaki and Karydis, Evangelos and Nishimura, Jun and Tsuchiya, Asato and Yamamori, Naoyuki",
    title = "{work in progress}",
}

@article{2209_10494,
    author = "Kumar, Arpith and Joseph, Anosh and Kumar, Piyush",
    title = "{Complex Langevin Study of Spontaneous Symmetry Breaking in IKKT Matrix Model}",
    eprint = "2209.10494",
    archivePrefix = "arXiv",
    primaryClass = "hep-lat",
    doi = "10.22323/1.430.0213",
    journal = "PoS",
    volume = "LATTICE2022",
    pages = "213",
    year = "2023"
}

@article{2308_03607,
    author = "Kumar, Arpith and Joseph, Anosh and Kumar, Piyush",
    title = "{Investigating Spontaneous SO(10) Symmetry Breaking in~Type IIB Matrix Model}",
    eprint = "2308.03607",
    archivePrefix = "arXiv",
    primaryClass = "hep-lat",
    doi = "10.1007/978-981-97-0289-3_337",
    journal = "Springer Proc. Phys.",
    volume = "304",
    pages = "1201--1203",
    year = "2024"
}

@article{1001_2933,
    author = "Witten, Edward",
    editor = "Andersen, Joergen E. and Boden, Hans U. and Hahn, Atle and Himpel, Benjamin",
    title = "{Analytic Continuation Of Chern-Simons Theory}",
    eprint = "1001.2933",
    archivePrefix = "arXiv",
    primaryClass = "hep-th",
    journal = "AMS/IP Stud. Adv. Math.",
    volume = "50",
    pages = "347--446",
    year = "2011"
}

@article{2501_17798,
    author = "Chou, Chien-Yu and Nishimura, Jun and Tripathi, Ashutosh",
    title = "{Inequivalence between the Euclidean and Lorentzian Versions of the Type IIB Matrix Model from Lefschetz Thimble Calculations}",
    eprint = "2501.17798",
    archivePrefix = "arXiv",
    primaryClass = "hep-th",
    reportNumber = "KEK-TH-2686",
    doi = "10.1103/PhysRevLett.134.211601",
    journal = "Phys. Rev. Lett.",
    volume = "134",
    number = "21",
    pages = "211601",
    year = "2025"
}

@article{0912_3360,
    author = "Aarts, Gert and Seiler, Erhard and Stamatescu, Ion-Olimpiu",
    title = "{The Complex Langevin method: When can it be trusted?}",
    eprint = "0912.3360",
    archivePrefix = "arXiv",
    primaryClass = "hep-lat",
    doi = "10.1103/PhysRevD.81.054508",
    journal = "Phys. Rev. D",
    volume = "81",
    pages = "054508",
    year = "2010"
}

@article{2507_18472,
    author = "Chou, Chien-Yu and Nishimura, Jun and Wang, Cheng-Tsung",
    title = "{Monte~Carlo Studies of the Emergent Spacetime in the Polarized Type IIB Matrix Model}",
    eprint = "2507.18472",
    archivePrefix = "arXiv",
    primaryClass = "hep-th",
    reportNumber = "KEK-TH-2740",
    doi = "10.1103/y1rm-n85b",
    journal = "Phys. Rev. Lett.",
    volume = "135",
    number = "22",
    pages = "221601",
    year = "2025"
}

@article{2604_19836,
    author = "Anagnostopoulos, Konstantinos N. and Azuma, Takehiro and Hirasawa, Mitsuaki and Nishimura, Jun and Papadoudis, Stratos and Tsuchiya, Asato",
    title = "{The emergence of (3+1)-dimensional expanding spacetime from complex Langevin simulations of the Lorentzian type IIB matrix model with deformations}",
    eprint = "2604.19836",
    archivePrefix = "arXiv",
    primaryClass = "hep-th",
    reportNumber = "KEK-TH-2826",
    month = "4",
    year = "2026"
}

\end{document}